
\magnification = 1200
\def\lapp{\hbox{$ {
\lower.40ex\hbox{$<$}
\atop \raise.20ex\hbox{$\sim$}
}
$}  }
\def\rapp{\hbox{$ {
\lower.40ex\hbox{$>$}
\atop \raise.20ex\hbox{$\sim$}
}
$}  }
\def\barre#1{{\not\mathrel #1}}
\def\krig#1{\vbox{\ialign{\hfil##\hfil\crcr
$\raise0.3pt\hbox{$\scriptstyle \circ$}$\crcr\noalign
{\kern-0.02pt\nointerlineskip}
$\displaystyle{#1}$\crcr}}}
\def\upar#1{\vbox{\ialign{\hfil##\hfil\crcr
$\raise0.3pt\hbox{$\scriptstyle \leftrightarrow$}$\crcr\noalign
{\kern-0.02pt\nointerlineskip}
$\displaystyle{#1}$\crcr}}}
\def\ular#1{\vbox{\ialign{\hfil##\hfil\crcr
$\raise0.3pt\hbox{$\scriptstyle \leftarrow$}$\crcr\noalign
{\kern-0.02pt\nointerlineskip}
$\displaystyle{#1}$\crcr}}}

\def\svec#1{\skew{-2}\vec#1}
\def\Tr{\,{\rm Tr }\,}

\def\g5{\gamma_5}

\def\lp1{{\cal L}_{\pi N}^{(1)}}
\def\lp2{{\cal L}_{\pi N}^{(2)}}
\def\lp3{{\cal L}_{\pi N}^{(3)}}

\topskip=0.40truein
\leftskip=0.18truein
\vsize=8.8truein
\hsize=6.5truein
\tolerance 10000
\hfuzz=20pt

\baselineskip 14pt plus 1pt minus 1pt
\pageno=0
\centerline{\bf NEUTRAL PION PHOTOPRODUCTION OFF NUCLEONS REVISITED}
\vskip 48pt
\centerline{V. BERNARD$^{1,\ddagger}$, N. KAISER$^{2,*}$,
Ulf-G. MEI{\ss}NER$^{3,\dagger}$}
\vskip 16pt
\centerline{$^1${\it Centre de Recherches Nucl\'{e}aires et Universit\'{e}
Louis Pasteur de Strasbourg}}
\centerline{\it Physique Th\'{e}orique,
BP 20Cr, 67037 Strasbourg Cedex 2, France}
\vskip  4pt
\centerline{$^2${\it Physik Department T30,
Technische Universit\"at M\"unchen}}
\centerline{\it
James Franck Stra{\ss}e, D-85747 Garching, Germany}
\vskip  4pt
\centerline{$^3${\it Institut f\"ur Theoretische Kernphysik,
Universit\"at Bonn}}
\centerline{\it
Nussallee 14--16, D-53115 Bonn, Germany}
\vskip  4pt
\vskip 12pt
\centerline{email: {$^\ddagger$}bernard@crnhp4.in2p3.fr,
$^*$nkaiser@physik.tu-muenchen.de,}
\centerline{ $^\dagger$meissner@pythia.itkp.uni-bonn.de}
\vskip 1.0in
\centerline{\bf ABSTRACT}
\medskip
\noindent
We investigate threshold pion photoproduction in the framework of
heavy baryon chiral perturbation theory. We give the expansion of
the electric dipole amplitude $E_{0+}$ to three orders in $\mu$, the
ratio of the pion to nucleon mass, and show that it is slowly
converging. We argue that this observable is not a good testing
ground for the chiral dynamics of QCD. In contrast, we exhibit new and
fastly converging low--energy theorems in the P--waves which should be
used to constrain the data analysis. We also discuss the importance of
polarization observables to accurately pin down certain multipoles and
give predictions for the reaction $\gamma n \to \pi^0 n$.
\medskip
\vfill
\noindent CRN 94--62 \hfill

\noindent TK 94 18 \hfill November 1994
\vskip 12pt
\eject
\baselineskip 14pt plus 1pt minus 1pt
\noindent{\bf I. INTRODUCTION}
\bigskip
Over the last few years, much interest has been focused on pion photo-- and
electroproduction off nucleons. In particular, new accurate data for
the processes $\gamma p \to \pi^0 p$ and $\gamma^\star p \to \pi^0 p$ close to
production threshold have become available [1,2].
 These have led to many theoretical investigations. In
particular, in refs.[3] baryon chiral perturbation theory was used to give a
model--independent description of the pertinent differential cross sections,
amplitudes and so on. In these papers, the nucleons were treated as
fully relativistic fields which leads to complications in the power
counting underlying the effective field theory [4]. These can be
overcome  by a clever choice of the spin--1/2 fields being
velocity--dependent which allows to transform the baryon mass term
into a string of $1/m$ (here, $m$ denotes the nucleon mass) suppressed
interactions [5,6]. In this formulation, called heavy baryon chiral
perturbation theory (HBCHPT), there is a strict
correspondence between the loop expansion and the expansion in small
momenta and quark masses. Our motivation to come back to the topic of
threshold pion photoproduction off protons and neutrons is fivefold.
First, the calculation in the relativistic framework indicated that
the expansion of the electric dipole amplitude $E_{0+}$ is indeed
slowly converging as a function of $\mu = M_\pi / m$, with $M_\pi$
denoting the pion mass. However, in that approach the complete
expression of order $\mu^3$ could not be given. In HBCHPT, if one
calculates within the one--loop approximation but to
next--to--next--to--leading order $q^4$ ($q$ denotes any small momentum
or mass), the first three terms in the chiral expansion of $E_{0+}$
can be given. Second, in the relativistic formulation [3] we found
that the differential cross sections were not well described
due to an essentially energy--independent $E_{0+}$ of $-1.3 \cdot
10^{-3}/M_{\pi^+}$. Also, in the P--waves the relative strength of the
$M_{1-}$ multipole came out too large as compared to the $M_{1+}$.
Here, we will show that
two of the P--waves are severely constrained by novel low--energy
theorems and that the third one is completely dominated by the $\Delta
(1232)$ resonance. These LETs were implicitely contained in the relativistic
calculation but not made explicit. Furthermore, we give here a much
better estimation of the pertinent low--energy constants based on the
idea of resonance saturation.
This in turn leads to a satisfactory description of
the existing data in the threshold region. Third, it is obvious that
the information gained from the total and differential cross
sections is not sufficient to
 pin down all multipoles uniquely. For doing that,
one has to consider polarization observables and we will discuss some
of these here. Last, but not least, we also show detailed predictions
for the reaction $\gamma n \to \pi^0 n$ which is experimentally  very
difficult to access but  in fact has to be studied for various
reasons, one of them being the test of the isospin decomposition
based on first order electromagnetism  which is
usually assumed in the construction of the invariant amplitudes.
Finally, new data  for $\gamma p \to \pi^0 p$ in the threshold region
and above taken at MAMI (Mainz) and SAL (Saskatoon)
are presently being analyzed.
\medskip
The manuscript is organized as follows.
In section 2, we fix our notation and define the
pertinent observables to be discussed later.
In section 3, we briefly
discuss the effective pion--nucleon Lagrangian underlying our calculation.
Sections 4 and 5 contain the
main results of this paper. We present the order $q^4$
calculation for the S--wave
including some isospin--breaking from the pion mass difference and the
corresponding ${\cal O}(q^3)$ calculation for the
P--waves.\footnote{$^{1)}$}{Due to the fast convergence in these multipoles, a
more accurate calculation does not seem necessary.} We then discuss
the low--energy theorems (LETs) in the S-- and P--waves followed by the
presentation and discussion of the numerical results. There are two
low--energy constants entering the expression for $E_{0+} (\omega)$
(with $\omega$ the pion cms energy). These can either be fixed by a
best fit to the available data or estimated by resonance
exchange. Already here we would like to stress that the first method
leads to unnaturally large numbers for these coefficients which are a
reflection of the importance of higher loops not yet
calculated. Consequently, even to this order there remains some appreciable
theoretical uncertainty leading to the conclusion that the electric
dipole amplitude is in fact not the best testing ground of the chiral
dynamics of QCD as was long believed. We also present a two--parameter
model for $E_{0+} (\omega)$
 which simulates most of the physics in the threshold region.
Furthermore, the imaginary part of $E_{0+}$ and its relation to the
Watson final state theorem is discussed showing again that to this
order the description of the S--wave is not too accurate. However, we
also stress that the presently available determinations of $E_{0+}$
close to threshold hinge on a few empirical points because the
P--waves quickly dominate the cross section. Therefore, it is
imperative to study polarization observables. These allow for a clean
separation of certain multipoles and a much more accurate empirical
determination of small multipoles like e.g. $E_{1+}$. The summary and
outlook is given in section 6 and some lengthy formulae are collected
in the appendix.

\goodbreak \bigskip
\noindent{\bf II. THRESHOLD PION PHOTOPRODUCTION: FORMAL ASPECTS}
\bigskip
In this section, we will give the formalism necessary to treat pion
photoproduction in the threshold region. We will only be concerned with
the kinematics close to threshold and the corresponding multipoles. We
also summarize the formulae for the differential
and the total cross sections as well as for some polarization
observables.
\medskip
Consider the process $\gamma(k) + N(p_1) \to \pi^0(q) + N(p_2)$,
with $N$ denoting the nucleon (proton or neutron), $\gamma$ a real ($k^2= 0$)
photon and $\pi^0$ the neutral pion. The
polarization vector of the photon is
denoted by $\epsilon_\mu$. In the threshold region, the
three--momentum $\vec q$ of the pion in the $\pi N$ centre--of--mass
(cm) frame is small and vanishes at threshold. It is therefore
advantageous to perform a multipole decomposition since at threshold only the
S--wave survives and close to threshold one can confine oneself to
S-- and P--waves. The corresponding multipoles are called
$(E, \, M)_{l \pm}$, where $E, \,
M$ stands for electric and magnetic,
 $l = 0,1,2, \ldots$ the pion orbital angular momentum and
the $\pm$ refers to the total angular momentum of the pion-nucleon system, $j
=l \pm 1/2$. These multipoles parametrize the structure of the
nucleon as probed with low energy photons.
Consequently, the T--matrix $T \cdot \epsilon$
  depends on four multipoles and takes the following form in the cm system
$${m \over 4 \pi \sqrt{s} } \,
 T \cdot  \epsilon = i \vec \sigma \cdot \vec \epsilon \, (E_{0+}
+ \hat k \cdot \hat q P_1)
+i \vec\sigma \cdot \hat k \, \vec \epsilon \cdot \hat q \,P_2
+ (\hat q \times \hat k ) \cdot \vec \epsilon \, P_3
\eqno(2.1) $$
The quantities $P_{1,2,3}$ represent the following combinations of the three
$P$-waves, $E_{1+}, \, M_{1+}$ and  $M_{1-}$,
$$\eqalign{
P_1 &= 3E_{1+} + M_{1+} - M_{1-}  \cr
P_2 &= 3E_{1+} - M_{1+} + M_{1-}  \cr
P_3 &= 2 M_{1+} + M_{1-} \cr} \eqno(2.2) $$
These four amplitudes are  calculable within CHPT. As will
become clear later, the particular choice (2.2) of the P--waves is
best suited for the chiral expansion and the physics related to
it.

We now discuss briefly the kinematics for $\gamma p \to \pi^0 p$.
The pion energy in the cm system is given by
$$ \omega = { s - m^2_p + M_{\pi^0}^2 \over 2 \sqrt{s}}
= {E_\gamma + M_{\pi^0}^2 /2m_p \over \sqrt{1 + 2E_\gamma / m_p}}
  \eqno(2.3)$$
with $s$ the cms energy squared and $E_\gamma
= (s-m_p^2)/(2m_p)$ the photon energy in
the lab frame. At threshold, $s_{\rm thr} =
(m_p + M_{\pi^0})^2$ and $\omega_{\rm thr} =
\omega_0 =  M_{\pi^0}$. The second threshold
 is related to the opening of the $\pi^+ n$ channel at $\omega_c = 140.11$ MeV
(since $s_c = (m_n + M_{\pi^+})^2$). In what follows, we will take into account
the pion mass difference $M_{\pi^\pm} - M_{\pi^0} = 4.6$ MeV since it subsumes
the most important isospin--breaking effects. This is discussed in
some detail
 in ref.[7]. However, we do not differentiate between the proton and
the neutron mass in the loops.
Consequently,  we will use $M_{\pi^\pm} = \omega_c =140.11$ MeV
to account for the proper location of the second threshold. The tiny error
induced by
this procedure is well within the theoretical uncertainty of our approach.

The differential cross section can be written as
$$ {|\vec k \, | \over |\vec q \,|}{d \sigma \over d \Omega}_{\rm cm}
 = A + B \cos \theta + C \cos^2
\theta \eqno(2.4)$$
in the approximation that only S-- and P--waves contribute.
$\theta$ is the cms scattering angle,  $|\vec k \,
|= (s-m_p^2)/(2 \sqrt{s})$ and $|\vec q \, | =
\sqrt{\omega^2-M^2_{\pi^0}}$.
The energy--dependent coefficients $A,B$ and $C$ are related to the
multipoles via
$$\eqalign{ A & = |E_{0+}|^2 +  {1 \over 2} |P_2|^2 + {1 \over 2} |P_3|^2
\cr B & = 2 \, {\rm Re} \, (E_{0+} P_1^* ) \cr C & =   |P_1|^2 -
{1 \over 2} |P_2|^2 - {1 \over 2} |P_3|^2 \cr} \eqno(2.5)$$
These are real for $\omega_0 \le \omega \le \omega_c$ and complex above
$\omega_c$
(we do not consider here the tiny phase related to the direct $\pi^0
p$
scattering process [8] because it only shows up at two--loop accuracy).
The total cross section follows as
$$ {|\vec k \, | \over |\vec q \,|} \sigma_{\rm tot} = 4 \pi \biggl( A
+ {1 \over 3} C \biggr) \quad . \eqno(2.6)$$
 From the nearly forward--backward
symmetric angular distributions exhibited by the Mainz
data, one can immediately conclude that $|A|,|C| \gg |B|$ which means
that a very accurate knowledge of the P--waves is mandatory to
reliably extract the electric dipole amplitude. Therefore, different
assumptions on the P--waves can lead to a rather different energy
variation of the electric dipole amplitude
 $E_{0+} (\omega )$ in the threshold region [9]. Also, in
most analysis it is assumed that $E_{1+} = 0$. To get a handle on such
small multipoles and to allow for a clean separation of the various
real and imaginary parts, one has to investigate polarization
observables. We will consider here the polarized photon asymmetry
$\Sigma (\theta )$, the polarized target asymmetry $T(\theta )$ and
the recoil polarization $P(\theta )$. These are given by
$$\eqalign{ \Sigma (\theta ) & = \Gamma \,  \sin \theta \, ( |P_3|^2 -
|P_2|^2 ) \cr T(\theta) & = 2\Gamma \, {\rm Im} \, ((E_{0+} + \cos \theta
P_1)(P_3 - P_2)^* ) \cr P(\theta ) & = 2\Gamma \,
{\rm Im} \, ((E_{0+} + \cos \theta P_1)(P_2 + P_3)^* ) \cr} \eqno(2.7)$$
with
$$ \Gamma = {|\vec q \,| \sin \theta \over 2 |\vec k \, |}
 \biggl( {d\sigma \over d\Omega}
\biggr)^{-1}_{\rm cm}  \quad . \eqno(2.7a)$$
In fact, $P(\theta) $ and $T(\theta )$ allow for a direct
determination of Im $E_{0+}$ above $\omega_c$ since the P--waves are
essentially real in the threshold region (as we will show later). This
concludes the necessary formalism.
\goodbreak \bigskip
\noindent{\bf III. EFFECTIVE LAGRANGIAN}
\medskip
In this section, we will briefly discuss the chiral effective Lagrangian
underlying our calculation.
To explore in a systematic fashion the consequences of spontaneous and explicit
chiral symmetry breaking of QCD, we make use of
 baryon chiral perturbation theory
(in the heavy mass formulation) (HBCHPT). The nucleon mass is
considered large compared to typical momenta in the system.
This allows to decompose the nucleon Dirac spinor into "large"
$(H)$ and "small" ($h)$ components
$\Psi(x) = e^{-i m v \cdot x } \{ H(x) + h(x)\}$
with $v_\mu$ the nucleon four-velocity, $v^2 = 1$, and the velocity eigenfields
are defined via $ \barre v H = H$ and $\barre v h = - h$.
Eliminating the "small" component field $h$ (which generates $1/m$
corrections), the leading order chiral $\pi N$ Lagrangian reads
$${\cal L}_{\pi N}^{(1)} = {\bar H} ( i v\cdot D + g_A S \cdot u ) H
\eqno(3.1)$$
Here the pions are collected in a SU(2) matrix-valued field $U(x)$
$$ U(x) = {1\over F} \bigl[ \sqrt{ F^2 - {\svec \pi}(x)^2 } + i \svec \tau
\cdot \svec \pi(x) \bigr] \eqno(3.2)$$
with $F$ the pion decay constant in the chiral limit and the so-called
$\sigma$-model gauge has been chosen which is of particular convenience for our
calculations in the nucleon sector. In eq.(3.1) $D_\mu = \partial_\mu +
\Gamma_\mu$ denotes the nucleon chiral covariant derivative, $S_\mu$ is a
covariant generalization of the Pauli spin vector, $g_A \simeq 1.26$ the
nucleon axial vector coupling constant (formally the one in the chiral limit)
and
$u_\mu = i u^\dagger \nabla_\mu U u^\dagger$,
with $u = \sqrt{U}$ and $\nabla_\mu$ the covariant derivative acting on the
pion fields. To leading order, ${\cal O}(q)$ one has to calculate tree
diagrams from
$$ {\cal L}^{(1)}_{\pi N} + {F^2 \over 4}{\rm Tr}\bigl\{ \nabla^\mu U
\nabla_\mu U^\dagger + \chi_+ \bigr\}  \, \, \, ,
\quad \chi_\pm = u^\dagger \chi u^\dagger \pm u \chi^\dagger u
\eqno(3.3)$$
where the second term is the lowest order mesonic chiral effective Lagrangian,
the non-linear $\sigma$-model coupled to external sources. The scalar source
$\chi$ is proportional to the quark mass matrix ${\cal M}$.
Beyond leading order,
the effective Lagrangian takes the form
$${\cal L}_{\rm eff} = {\cal L}_{\pi N}^{(1)} + {\cal L}_{\pi N}^{(2)} +
 {\cal L}_{\pi N}^{(3)} +{\cal L}_{\pi N}^{(4)} + {\cal L}_{\pi \pi}^{(2,4)}
 \eqno(3.4)$$
${\cal L}_{\pi N}^{(2,3,4)}$ contain $1/m$ corrections and
counterterms. The a priori unknown coefficients of these counterterms
 are the so-called low energy constants. For the calculation to order
$q^4$, one has to consider tree diagrams with insertions from
${\cal L}_{\pi N}^{(2,3,4)}$ as well as one--loop diagrams with
insertions from ${\cal L}_{\pi N}^{(1,2)}$. The terms from
${\cal L}_{\pi N}^{(2)}$ of relevance to the problem at hand are
$$\eqalign{
{\cal L}_{\pi N}^{(2)} & =  {\bar H} \biggl\lbrace  -{1\over 2m} D \cdot D +
{1 \over 2m} (v \cdot D)^2
- {i g_A \over 2m} \lbrace S \cdot D , v \cdot u
\rbrace \cr & - { i\over 4 m}[S^\mu , S^\nu ]
\bigl( (1 + \krig\kappa_v) f^+_{\mu
\nu} + {\krig\kappa_s - \krig\kappa_v  \over 2 } \Tr f^+_{\mu \nu} \bigr)
\biggr\rbrace H \cr} \eqno(3.5)$$
where $f_{+ \mu \nu} = e (u^\dagger Q u + u Q u^\dagger)  F_{\mu \nu}$,
$Q = (1 +\tau_3)/2$ and $F_{\mu \nu}$ is the canonical photon field
strength tensor. Note that ${\krig \kappa}_{S,V}$, the isoscalar and isovector
anomalous magnetic moments of the nucleon in the chiral limit, are
combinations of the low--energy constants $c_6$ and $c_7$ discussed in
[7]. We note here that the other low--energy constants of order $q^2$,
which are called $c_{1,2,3,4}$, do not contribute at all to our final
results for the chiral expansion of the S-- and P--waves (to the order we
are working).
There are two terms from ${\cal L}_{\pi N}^{(3)}$ which
enter [10],
$${\cal L}_{\pi N}^{(3)} =
\bar H \biggl\lbrace b_{14} \, v_\lambda
\epsilon^{\lambda \mu \nu \rho} \Tr (f_{+ \mu \nu} u_\rho ) + b_{22} \,(
i v_\lambda \epsilon^{\lambda \mu \nu \rho} \, S_\rho \, f_{+ \mu \nu}^V \, v
\cdot D + {\rm h.c.}) \biggr\rbrace H + \dots
\eqno(3.6)$$
with $f_{+ \mu \nu}^V = f_{+ \mu \nu} - (1/2) \,{\rm Tr}f_{+ \mu \nu}$.
Although the coefficients $b_{14}$ and $b_{22}$ are infinite [10], in the
case at hand only the finite combination
$ b_P \sim 4 \, b_{14} + \krig{g}_A \, b_{22}$
contributes and the low--energy constant $b_P$ enters the P--wave
$P_3$ in the following way:
$$P_3^{\rm ct}(\omega ) = e \, b_P \, \omega \,|\vec q \,| \quad . \eqno(3.7)$$
Note that for the other multipoles $E_{0+}$ and $P_{1,2}$ one can not
construct any gauge and Lorentz invariant chirally symmetric counter
term at ${\cal O}(q^3)$ respecting the discrete symmetries P, C, and T.
 Finally, the minimal set of terms from ${\cal L}_{\pi N}^{(4)}$
which give a contribution to $E_{0+} (\omega )$ (for the proton) is of the form
$${\cal L}_{\pi N}^{(4)} =
\bar H \biggl\lbrace -4 \pi F \, a_2
 \, S^\mu \, v^\nu \, f_{\mu \nu}^+
\, ( v \cdot D v \cdot u ) \, - \, 2 \pi i F \, a_1 \, S^\mu \,
v^\nu \, f_{\mu \nu}^+ \chi_- \biggr\rbrace H + \dots
\eqno(3.8)$$
and show up in $E_{0+}$ in the following way,
$$E_{0+}^{\rm ct} (\omega ) = e \, a_1 (\lambda) \omega \, M_\pi^2 \, + \, e \,
a_2 (\lambda) \, \omega^3 \, \quad . \eqno(3.9)$$
These counter terms are necessary to absorb the divergences generated
by the loops at order $q^4$.
Here, $\lambda$ is the scale introduced via the dimensional
regularization. Ideally, this scale dependence is cancelled from the one
in the corresponding loop contribution. If one, however, estimates the
contact terms via resonance saturation, one is left with a small but spurious
scale dependence as detailed in ref.[13]. We also note that the
low--energy constants $b_P$, $a_1$ and $a_2$ are a priori different for the
proton and the neutron. The dots in eqs.(3.6,3.8) stand for the
respective chiral vertices accompanied by the low--energy constants.
We omit the superscript '(p,n)' here since
from the context it is obvious which reaction we are considering.
There are furthermore many terms in ${\cal L}_{\pi N}^{(3,4)}$ which
come from the $1/m$--expansion of the relativistic $\pi N$ Dirac
Lagrangian. These generate the $1/m$--expanded Born contributions to
the S-- and P--waves given in the appendix.
Finally, we note that we will work in the Coulomb gauge $\epsilon_0 =0$
since in that gauge one has not direct lowest--order photon--nucleon
coupling (which is proportional to $\epsilon \cdot v$) and thus many
diagrams vanish.
\goodbreak \bigskip
\noindent{\bf IV. QCD ANALYSIS}
\medskip
In this section, we briefly outline the ${\cal O}(q^3)$ calculations for the
P--waves and the ${\cal O}(q^4)$ for $E_{0+}$. The explicit
expressions are relegated to the appendix. Next, we estimate the
low--energy constants which enter the expressions for $E_{0+}$ and $P_3$.
We then turn to the discussion of
the low--energy theorem (LET) for $E_{0+}$ and exhibit novel LETs for the
P--waves. This constitutes one of the major results of this investigation.
We end this section with a short discussion of the imaginary part of
$E_{0+}$ and its relation to the Fermi--Watson theorem.
\goodbreak \bigskip
\noindent{\bf IV.1. CHIRAL EXPANSION TO ORDER ${\cal O}(q^3)$}
\medskip
We wish to calculate the T--matrix elements to order $q^3$. Since the photon
polarization vector counts as ${\cal O}(q)$, one gets the S-- and P--wave
multipoles with an accuracy of ${\cal O}(q^2)$. In the Coulomb gauge, we have
tree contributions from ${\cal L}_{\pi N}^{(2,3)}$ to $E_{0+}$ and $P_i$
$(i=1,2,3)$ of the type
$$\eqalign{ E_{0+} & \sim {\omega \over m^2}
f_2( {\omega \over M_\pi} )\, , \, \,
{\omega^2 \over m^3} f_3( {\omega \over M_\pi} )
\cr P_i & \sim {|\vec q \,| \over m^2} \, , \, \,
 |\vec q \,| {\omega \over m^3} g_i( {\omega \over
M_\pi} )  \cr} \eqno(4.1) $$
where $f_2, \, f_3$ and $g_i$ are dimensionless functions of their
arguments. In the one--loop diagrams,
we include the pion mass difference to account for the most important
isospin--breaking effect. $E_{0+} (\omega)$
 is given by the triangle and rescattering
diagrams (see fig.1a,b) and the other two diagrams shown in that figure
contribute to $P_{1,2} (\omega )$. To this order, there is no loop contribution
to $P_3 (\omega )$. The pion mass difference in the loops leads to the
cusp in  $E_{0+}$ of the square--root type
$$  E_{0+} (\omega )= {\rm coefficient} \cdot
 \sqrt{ 1 - {\omega^2 \over \omega_c^2}} + \ldots    \eqno(4.2)$$
where the ellipsis stands for polynomial pieces (in $\omega$). We will come
back to the coefficient multiplying the square root when we discuss the
final state theorem. At this order, there is no mass and coupling constant
renormalization due to the vanishing of the tree couplings.
Finally, there are also finite counter terms at this order.
The anomalous magnetic moment of the proton (neutron) enters the expressions
for $E_{0+}$ and $P_{1,2}$. In addition, there is the finite coefficient $b_P$
defined in eq.(3.7) contributing to $P_3$. The estimate of its numerical value
will be given in section 4.3.
\goodbreak \bigskip
\noindent{\bf IV.2. CHIRAL EXPANSION TO ORDER ${\cal O}(q^4)$}
\medskip
For the electric dipole amplitude, we consider one more order. This is
motivated by the fact that the relativistic calculation of ref.[3] points
towards the importance of higher orders. On more general grounds, we remind the
reader of the well--known fact that in S--wave observables it is often
mandatory to go beyond ${\cal O}(q^3)$. However, we should also stress already
at this point that a calculation up--to--and--including ${\cal O}(q^4)$ might
not be sufficiently accurate.  Furthermore, we do not differentiate here
between the neutral and charged pion masses
and use only $M_{\pi^+}$ to get the proper cusp effect at this order.
In the polynomial pieces (of the loops), this inflicts a
theoretical error proportional to $(\omega - M_{\pi^+}) / M_{\pi^+}
\sim 4 \%$  which is smaller than
the uncertainty from the determination of the low--energy
constants. The dominant unitarity (cusp) effect is already accounted
for at order $q^3$, compare eq.(4.2). At order $q^4$,
 there are tree diagrams which lead
to terms of the type $\omega^3 / m^4$ and there are {\it many} loop graphs. The
latter can be categorized as follows: (i) loop graphs with one vertex from
${\cal L}_{\pi N}^{(1)}$ and one vertex from ${\cal L}_{\pi N}^{(2)}$, (ii)
with vertices from ${\cal L}_{\pi N}^{(1)}$ but a nucleon propagator from
${\cal L}_{\pi N}^{(2)}$  and (iii) the $1/m$ corrections to the loops
calculated at order $q^3$. Of course, many of the loop diagrams account for
mass and coupling constant renormalization,
$$( \krig{m} , \krig{g}_{\pi N} , \krig{\kappa}_{p,n}, F , M ) \to
(m , g_{\pi N}, \kappa_{p,n} , F_\pi , M_\pi )  \quad . \eqno(4.3)$$
Finally, there are the novel counterterms which contribute to $E_{0+}$ as given
in eq.(3.9). Therefore, the  expressions for $E_{0+}$ and $P_{1,2,3}$
calculated to order $q^3$ and $q^4$, respectively, take the generic form
$$\eqalign{ E_{0+}(\omega) & = E_{0+}^{\rm Born}(\omega) +
E_{0+}^{\, \rm q^3-loop}(\omega) + E_{0+}^{\, \rm q^4-loop}(\omega) +
E_{0+}^{\rm ct}(\omega) \cr
P_i (\omega ) & = P_i^{\rm Born}(\omega) + P_i^{\, \rm q^3-loop}(\omega)\, ,
\quad i=1,2 \cr
P_3 (\omega ) & = P_3^{\rm Born}(\omega) + P_3^{\rm ct}(\omega) \cr}
 \eqno(4.4) $$
where 'Born' subsumes the nucleon--pole and anomalous magnetic moment
contributions.\footnote{$^{2)}$}{This
decomposition facilitates the comparison with
the existing literature but is not a consequence of the chiral expansion.}
The explicit expressions for
the various terms appearing in eq.(4.4) can be found in the
appendix. Here, we just remark that the Born and counterterm contributions are
real and that the loop contributions are complex for $\omega > \omega_c$. We
also point out again that the leading terms for $E_{0+}$ and $P_{1,2}$
appear at  the same
chiral power, namely ${\cal O}(q^2)$.
Indeed, the P--waves are proportional to $|\vec q \,|$, but not
to $|\vec q \,|\, |\vec k \,|$ as usually assumed. However, since $|\vec k \,|$
only varies by four per cent from
$\omega_{\rm thr}$ to $\omega_c$, this makes no
visible effect. Note furthermore that $P_3$ is essentially given by the contact
term $P_3^{\rm ct}$ since the Born contribution to this
 multipole is very small. It is also instructive to notice that the
magnetic part of the nucleon Born terms  is dominated by the
$M_{1+}$ and $M_{1-}$ multipoles with $2M_{1+} + M_{1-} \simeq 0$
 whereas static $\Delta$ exchange leads
approximatively to $P_1 = -P_2 \simeq P_3 /4$, i.e. $2M_{1+} + M_{1-}$
is much  larger than $M_{1+} - M_{1-}$. This is, of course, a particular
feature of the threshold region, further up in energy the $M_{1+}$
multipole quickly becomes dominant.
\goodbreak \bigskip
\noindent{\bf IV.3. ESTIMATION OF THE LOW--ENERGY CONSTANTS}
\medskip
The most difficult task is to pin down the values of the low--energy
constants (LECs) $a_1$, $a_2$ and $b_P$. We concentrate on the
reaction $\gamma p \to \pi^0 p$ and mention the necessary
modifications for the case $\gamma n \to \pi^0 n$ in the end of this
section.  We will follow two approaches
here. In the first one, we will use the available total and
differential cross section data to fix these coefficients.
However, as we will see, this leads to unnaturally large values of the
constants $a_1$ and $a_2$ because they subsume the effects of higher
loops not yet calculated. Therefore and secondly, we will
 estimate the LECs making use of the resonance saturation principle
[11]. This can be formulated as follows. Consider meson resonances (M=V,A,S,P)
and baryonic excitations ($N^\star = \Delta(1232), N^*(1440), \ldots$)
chirally coupled to the Goldstone bosons (collected in $U$) and the
matter fields ($N$). Integrating out the meson and
nucleon excitations,
$$ \int [dM][dN^\star] \exp \, i \int dx \,
\tilde{{\cal L}}_{\rm eff} [U,M,N,N^\star] =
\exp \,i \int dx \, {\cal L}_{\rm eff} [U,N] \eqno(4.5)$$
one is left with a string of higher dimensional operators contributing
to ${\cal L}_{\rm eff} [U,N]$ in a manifestly chirally invariant manner
and with coefficients given entirely in terms of resonance masses and
coupling constants of these resonance fields to the Goldstone
bosons. A specific example for the baryon sector is discussed in
ref.[12]. For the case at hand, we have mesonic and baryonic
contributions. As already discussed in [3], there is t--channel vector
meson exchange, here the $\rho (770)$ and the $\omega(782)$. These
contribute to $a_{1,2}$ as follows ($V = \rho^0 + \omega$),
$$a_2^V = { 5 \over 48 \pi^3 m F_\pi^3} = 4.45 \, {\rm GeV}^{-4}
\, , \quad a_1^V = -{2 \over 5} a_2^V = -1.78 \, {\rm GeV}^{-4}
 \eqno(4.6)$$
where we have used $g_{\rho N} = g/2$, $g_{\omega N} = 3g/2$, $M_\rho
= M_\omega = \sqrt{2} g F_\pi$, $g=5.8545$, $\kappa_\rho = 6$,
$\kappa_\omega = 0$ and $G_{\pi \rho \gamma} = g / (16 \pi^2 F_\pi) =
(1/3) G_{\pi \omega \gamma}$ together with $m =938.28$ MeV and $F_\pi
= 93$ MeV.  Notice that the values of these coupling constants are somewhat
simplified and given in part by the gauged Wess--Zumino
action. However, as it will turn out, the vector meson contribution to
the P--wave $P_3$ is rather small and in the S--wave, there are
uncertainties due to higher order effects so that the
accuracy given by these values is sufficient. We could as well take
the vector dominance value of $\kappa_\omega = -0.12$ (and so on) without any
noticeable change in the numerical results to be discussed later.
The sign of the vector meson contribution is fixed
from the sign of the triangle anomaly. The P--wave
contribution is found as
$$b_P^V = {5 \over 64 \pi^3 F_\pi^3} =
3.13 \, {\rm GeV}^{-3}  \quad . \eqno(4.7)$$
 The result given in eq.(4.7) is not affected by the tensor coupling
in agreement with the considerations presented in ref.[3].

 From the baryon sector, the by far largest contribution comes from the
$\Delta(1232)$ resonance. It is mandatory to consider the $\Delta$ as
a fully relativistic spin--3/2 field before integrating it
out. Therefore, the contribution to the various low--energy constants
will depend on the two $\Delta N \gamma$ coupling constants $g_1$ and
$g_2$, the $\pi N \Delta $ coupling $g_{\pi N \Delta} = (3/\sqrt{2}) g_{\pi
N}$ and the off--shell parameters $X$, $Y$ and $Z$ (see also the
discussion in ref.[13]). This procedure leads to
$$\eqalign{
a_1^\Delta & = {C g_1 \over 6} \biggl[ {-2m_\Delta^2 + 2m_\Delta
m + m^2 \over m_\Delta - m} + 2m (2Y - Z - 2YZ) - 2 m_\Delta (Y + Z
+ 4YZ)\biggr] \cr & +
{C g_2 \over 8} \biggl[ m(2X+1) + m_\Delta
\biggr] \cr}    \eqno(4.8a)$$
$$\eqalign{
a_2^\Delta & = {C g_1 \over 12} \biggl[ {10m_\Delta^2 - 7m_\Delta
m - 8m^2 \over m_\Delta - m} + 2m (5Y + 2Z - 2YZ) + 10 m_\Delta (Y + Z
+ 4YZ)\biggr] \cr & +
{C g_2 \over 16} \biggl[ m(2X+1)(1-6Z) + 2m_\Delta
(4XZ+X+Z) \biggr] \cr}    \eqno(4.8b)$$
with $C = g_{\pi N} / (6 \sqrt{2} \pi m^3 m_\Delta^2 ) = 0.40$
GeV$^{-5}$ and $g_{\pi N} =13.4$. Throughout, we use the
Goldberger--Treiman relation to fix $g_A$, i.e. $g_A = g_{\pi N} F_\pi
/ m$.
In what follows, we will adopt two strategies. First, we keep $g_1 =
g_2 =5 $ fixed and vary $X,Y,Z$ in the ranges given in ref.[14] and
then also allow to vary $g_1$ and $g_2$ within the ranges given in
[14]. As it will turn out, the results are fairly insensitive to the
variation of $g_1$ and $g_2$. Inspection of eqs.(4.8a,b) reveals a
rather large uncertainty inflicted from the relatively poor knowledge
of the off--shell parameters. We will come back to this when we
discuss the numerical results. Furthermore, the $\Delta$ contributes
prominently to $b_P$,
$$
b_P^\Delta  = {C g_1 m \over 2} \biggl[ {2m_\Delta^2 + m_\Delta
m - m^2 \over m_\Delta - m} + 2m (Y + Z + 2YZ) + 2 m_\Delta (Y + Z
+ 4YZ)\biggr]
  \eqno(4.9)$$
Notice that to the order we are working, $b_P^\Delta$ receives no
contribution proportional to $g_2$.
To get an idea about the size of $b_P^\Delta $, we use the static
isobar model (which involves no off--shell parameters) and find
$$b_P^{\Delta, {\rm static}}  = { g_1  g_{\pi N} (m_\Delta - m)
\over 6 \sqrt{2} \pi m^2 ((m_\Delta -m)^2 - M_\pi^2)} = 12.3 \, {\rm GeV}^{-3}
\eqno(4.10)$$
which is considerably larger than the vector meson contribution, eq.(4.8).
For the neutron, the only difference is the sign of the $\rho^0$--meson
contribution. We get
$$ a_i^{V,n} = {1 \over 8} a_i^{V,p} \,\,  (i=1,2) \, ; \quad b_P^{V,n} =
{ 4 \over 5} \, b_P^{V,p}  \, \, .   \eqno(4.11)$$
In this case, we have no data to fit these LECs and must use the
resonance exchange estimates.
\goodbreak \bigskip
\noindent{\bf IV.4. LOW--ENERGY THEOREMS (LETs) FOR THE S-- AND P--WAVES}
\medskip
We consider here the expansion  in $\mu =M_\pi /m$
of $E_{0+}$, $P_1$ and $P_2$ at threshold. Since $P_3$ is completely
dominated by the contact term proportional to $b_P$, a similar
expansion for this combination of the P--wave multipoles does not make
sense. First, however, let us briefly state what is meant by a LET
following ref.[15]. We consider as a LET of order $q^n$ a
general prediction of CHPT to ${\cal O}(q^n)$. General prediction
means a strict consequence of the Standard Model depending on some
low--energy constants like $F_\pi, m, g_A, \kappa_p, \ldots$, but
without any model assumption for these parameters. This gives a
precise prescription for obtaining higher--order corrections to the
leading order LETs which can e.g. be obtained from current algebra.

First, we study the electric dipole amplitude. For that, we work in
the isospin limit $m_u = m_d$ and to first order in the
electromagnetic coupling constant. The $\mu$--expansion
takes the form
$$ E_{0+, {\rm thr}}^i = -{e g_{\pi N} \over 8 \pi m} \mu \lbrace \,
C_1^i + \,  \mu \, \, C_2^i +\, \mu^2 \, C_3^i \,
+ {\cal O}(\mu^3) \, \rbrace \, , \quad i=p,n
\eqno(4.12)$$
with
$$\eqalign{ C_1^p & = 1, \quad C_1^n = 0 \cr
C_2^p & = -{1 \over 2} ( 3 + \kappa_p +{m^2 \over 8 F_\pi^2}) \, , \quad
C_2^n = {1 \over 2} ( \kappa_n - {m^2 \over 8 F_\pi^2}) \cr
C_3^p & = {3 \over 4} ( {5 \over 2} + \kappa_p ) - {m^2 \over 16
\pi^2 F_\pi^2 } \biggl[ ( 10 + {8 \over 3} g_A^2 ) \ln {M_\pi \over
\lambda } - g_A^2 ( {\pi^2 \over 4} - {5 \pi \over 3} + {11 \over 9})
-4 ( 1 + {5 \pi^2 \over 16}) \biggr] \cr & - {8 \pi m^4 \over g_{\pi N}}(a_1^p
(\lambda) +a_2^p (\lambda))    \cr
C_3^n & = -{3 \over 4}  \kappa_n - {m^2 \over 16
\pi^2 F_\pi^2 } \biggl[{2 \over 3} ( 7 + 4 g_A^2 ) \ln {M_\pi \over
\lambda } - g_A^2 ( {\pi^2 \over 4} - {5 \pi \over 3} + {11 \over 9})
+ ( {4 \over 9} - {5 \pi^2 \over 4}) \biggr] \cr
& - {8 \pi m^4 \over g_{\pi N}}(a_1^n (\lambda) +a_2^n (\lambda))    \cr}
\eqno(4.12a)$$
At present, the LETs given by (4.12) do not have too much predictive
power since the only way to determine the LECs $a_1 (\lambda)$
and $a_2 (\lambda)$ are the threshold data from Mainz [1] (for the proton).
Using the best fit values (see section 5), one finds $a_1^p (m) +a_2^p
(m) = 2.7$ GeV$^{-4}$ leading to\footnote{$^{3)}$}{Notice that
the individual values
of $a_1^p (m)$ and $a_2^p (m)$ are much larger than their sum
so that this determination is afflicted with a substantial
uncertainty. Even the sum  $a_1 + a_2$ is strongly affected if one chooses
either the charged or the neutral pion mass in eq.(3.9).}
$$ E_{0+, {\rm thr}}^p = -3.45 \, ( \, 1 \, - \, 1.26 \, + \, 0.55 \,
+ \dots) \cdot 10^{-3} / M_{\pi^+} \, = \, -1.0 \cdot 10^{-3} / M_{\pi^+}
\eqno(4.13)$$
where the ellipsis stands for terms of order $\mu^4$ and higher.
 Setting $a_1^p (m) + a_2^p (m) = 0$, the 0.55 would read 0.64
and the corresponding value for $ E_{0+, {\rm thr}}^p = -1.33
\cdot 10^{-3} / M_{\pi^+}$. For the neutron, we find (using the same
values for the LECs)
$ E_{0+, {\rm thr}}^n = 3.64 \, ( \, 1 \, - \, 0.29
+ \dots) \cdot 10^{-3} / M_{\pi^+} \, = \, 2.59 \cdot 10^{-3} / M_{\pi^+}$
which shows a better convergence since the term of order $\mu$ is
absent and thus the contribution from the triangle diagram appears
already at lowest order. We note that the electric dipole amplitude
for $\pi^0$ production off neutrons is sizeable and of opposite sign
to the one for production off protons. The lesson to be
learned is that the $\mu$
expansion of $E_{0+}$ converges very slowly (as already anticipated in
ref.[3]) and thus one has at least to go one order higher before one
can make a reasonably accurate theoretical prediction. This clearly
shows that the electric dipole amplitude is {\it not} a good testing
ground for the chiral dynamics of QCD. Such a behaviour is, however,
not too surprising. We remind the reader of similar large higher order
S--wave effects in the scalar form factor of the nucleon [16] or the
scalar form factor of the pion [17] (just to name two such cases). To
further tighten the prediction on $E_{0+}$, it is conceivable that one
has to perform a dispersive analysis supplemented with CHPT
constraints to get a handle on the higher orders. It is again
important to stress that the LETs for the electric dipole amplitude
have been derived in the exact isospin limit. It is not
meaningful to compare the number (4.13) with the data,
since isospin breaking and other higher order effects are substantial.
\medskip
We now turn to the P--waves. From the explicit formulae given in the appendix
we  derive the LETs for the slope of $P_1$ and $P_2$
at threshold (for the proton)
$$\eqalign{
{1 \over |\vec q\,|} P_{1,{\rm thr}}
& = {e g_{\pi N} \over 8 \pi m^2}
\biggl\lbrace 1 + \kappa_p + \mu \biggl[ -1 - {\kappa_p \over 2} + {g_{\pi N}^2
(10 - 3\pi ) \over 48 \pi} \biggr] + {\cal O}(\mu^2) \biggr\rbrace \cr
{1 \over |\vec q\,|} P_{2,{\rm thr}}
& = {e g_{\pi N} \over 8 \pi m^2}
\biggl\lbrace -1 - \kappa_p + {\mu \over 2} \biggl[ 3 + \kappa_p
 - {g_{\pi N}^2  \over 12 \pi} \biggr]
+ {\cal O}(\mu^2) \biggr\rbrace \cr}  \eqno(4.14a)$$
with $\kappa_p = 1.793$ and similarly for the reaction $\gamma n \to \pi^0 n$,
$$\eqalign{
{1 \over |\vec q\,|} P_{1,{\rm thr}} & = {e g_{\pi N} \over 8 \pi m^2}
\biggl\lbrace - \kappa_n + {\mu \over 2}
 \biggl[ \kappa_n + {g_{\pi N}^2
(10 - 3\pi ) \over 24 \pi} \biggr] + {\cal O}(\mu^2)  \biggr\rbrace \cr
{1 \over |\vec q\,|} P_{2,{\rm thr}}
& = {e g_{\pi N} \over 8 \pi m^2}
\biggl\lbrace  \kappa_n - {\mu \over 2} \biggl[ \kappa_n
 + {g_{\pi N}^2  \over 12 \pi} \biggr]  + {\cal O}(\mu^2)
\biggr\rbrace \cr}   \eqno(4.14b)$$
with $\kappa_n = -1.913$ the anomalous magnetic moment of the neutron.
 These are examples of  quickly converging $\mu$
expansions,\footnote{$^{4)}$}{To be precise, we mean that the leading
term is much bigger than the first correction in contrast to what
happens in the electric dipole amplitude.}
$$\eqalign{
{1 \over |\vec q\,|} P_{1,{\rm thr}}^p & = 0.512 \, ( 1 - 0.062 )
 \, {\rm GeV}^{-2} = 0.480 \, {\rm GeV}^{-2} \cr
{1 \over |\vec q\,|} P_{2,{\rm thr}}^p & = -0.512 \, ( 1 - 0.0008 )
 \, {\rm GeV}^{-2} = -0.512 \, {\rm GeV}^{-2} \cr
{1 \over |\vec q\,|} P_{1,{\rm thr}}^n & = 0.351 \, (1 - 0.020)
 \, {\rm GeV}^{-2} = 0.344 \, {\rm GeV}^{-2} \cr
{1 \over |\vec q\,|} P_{2,{\rm thr}}^n & = -0.351 \, ( 1 + 0.107 )
 \, {\rm GeV}^{-2} = -0.389 \, {\rm GeV}^{-2} \cr}
\eqno(4.15)$$
 From these expressions, a few interesting observations can be
made. First, we note that the multipole $E_{1+}$ is not exactly zero
since in that case one would have $P_1 = -P_2$. Commonly, this
multipole is set to zero when one analyzes the threshold data. We will
come back to this small multipole when we discuss the polarization
observables. Second, these P--wave LETs help to constrain the existing
fits to the threshold region [9]. They favor the solution which leads
to a strong energy--dependence of the electric dipole amplitude. If we
pull out by hand a factor $|\vec k \,|$\footnote{$^{5)}$}{We stress
again that this is not the correct threshold behaviour of the P--wave
multipoles.}, these LETs translate into
$$P_{1,{\rm thr}}^p  = 10.3 \, |\vec k \,| |\vec q \,| 10^{-3} /
M_{\pi^+}^3 \,  \quad P_{2,{\rm thr}}^p  = -10.9 \, |\vec k \,| |\vec q
\,| 10^{-3} / M_{\pi^+}^3 \, \eqno(4.16)$$
to be compared e.g. with the value of $P_1 =(8.8 \pm 0.6) |\vec k \,| |\vec
q \,| 10^{-3} / M_{\pi^+}^3$ given by Drechsel and Tiator [18].
Third, we also would like to stress that the corresponding $P_1$ and
$P_2$ of the relativistic calculation [3] agree quite nicely with the LET
(remember that in the
relativistic formulation some higher order terms are included). For example,
at $E_\gamma = 151$ MeV, the LET predicts $P_1 = 2.47$ and $P_2 = -2.48$ while
the P--wave multipoles of ref.[3] lead to $P_1 = 2.43$ and $P_2 = -2.60$
(all in units of $10^{-3} / M_{\pi^+}$) (for $\gamma p \to \pi^0 p$).
These novel P--wave LETs should be tested more accurately and they can
also serve to constrain the empirical analysis. It is amusing to note
that this is a {\it good} testing ground for chiral dynamics in contrast to
common folklore.
\goodbreak \bigskip
\noindent{\bf IV.5. RELATION TO THE FERMI--WATSON THEOREM}
\medskip
Here, we wish to elaborate briefly on the imaginary part of the
electric dipole amplitude. By virtue of  the Fermi--Watson theorem, it is
related to the $\pi N$ scattering phases via (see e.g. [20])
$$ {\rm Im} \, E_{0+}^{\pi^0 p} = {\rm Re} \, E_{0+}^{(0)} \tan(\delta_1) +
{1 \over 3} {\rm Re} \, E_{0+}^{(1/2)} \tan(\delta_1) +
{2 \over 3} {\rm Re} \, E_{0+}^{(3/2)} \tan(\delta_3)    \eqno(4.17)$$
with $\delta_{1,3}$ the $\pi N$ S--wave phases for total isospin $1/2$ and
$3/2$, respectively. Close to threshold, we can approximate the $\tan
( \delta_{2I})$ by $a_{2I} \,
|\vec q \,|$ (in the respective channels) and also
drop the term proportional to $a^+ \, {\rm Re}\,E_{0+}^{\pi^0 p}$
which is a factor of 200 smaller than $\sqrt{2} \, a^- \,
{\rm Re} \, E_{0+}^{\pi^+ n}$, i.e.
$$ {\rm Im} \, E_{0+}^{\pi^0 p} \simeq
\sqrt{2} \, a^- \, M_\pi \,{\rm Re} \, E_{0+}^{\pi^+ n} \, \sqrt{
{\omega^2 \over M_\pi^2} - 1} \quad . \eqno(4.18)$$
Therefore, the strength of the imaginary part of $E_{0+}$ in the
threshold region is governed by the product $\sqrt{2} \, a^- \, M_\pi
\, {\rm Re} \, E_{0+}^{\pi^+ n} = 3.7 \cdot 10^{-3} / M_{\pi^+}$.
This is in fact the coefficient which we did not write explicitely in
eq.(4.2).  The order $q^4$ calculation leads to
$$ {e g_{\pi N} M_\pi^2 \over 32 \pi^2 m F_\pi^2} \, \bigl( 1 - {5 \mu
\over 2} \bigr) = 2.7 \cdot 10^{-3} / M_{\pi^+} \, , \eqno(4.19)$$
which shows that the strength of Im~$E_{0+}$ is underestimated by
approximately 30$\%$. As already noted a couple of times, this
indicates that for an accurate description of the electric dipole
amplitude even in the threshold region one has to go beyond order
$q^4$. Of course, we should also stress that the imaginary part is in
any case less accurately calculated. Here, the first
contributions to Re~$E_{0+}$ are of order $q^2$ whereas the
corresponding imaginary part starts at ${\cal O}(q^3)$.
 Finally, we wish to point out why
the phase related to the direct $\pi^0 p$ scattering process only
appears at two--loop accuracy making use of the Fermi--Watson
theorem. Above the $\pi^0 p$ but below the $\pi^+ n$ threshold,
we have Im~$E_{0+}^{\pi^0 p}
= $ Re~$E_{0+}^{\pi^0 p} \cdot \tan \delta(\pi^0 p)$. In the
T--matrix, $a^+$ starts at order $q^2$ and Re~$E_{0+}^{\pi^0 p}$ is
${\cal O}(q)$. Furthermore, there is an additional factor $q$ from the
relation $\tan \delta^{\pi^0 p} = a^+ \, | \vec q \, |$
so that the imaginary part starts out at order
$q^4$, i.e. is ${\cal O}(q^5)$ in the full amplitude which is a two--loop
effect.
\goodbreak \bigskip
\noindent{\bf V. RESULTS AND DISCUSSION}
\medskip
In this section, we will discuss first a two--parameter model which
describes most of the physics in the threshold region and draw some
general conclusions from it. We then turn to the detailed numerical
investigation of the reaction $\gamma p \to \pi^0 p$ making use of the
existing Mainz data [1]. We will perform a free fit to the three
low--energy constants $a_1$, $a_2$ and $b_P$ but also one constrained
by resonance exchange considerations. We argue that the latter one is
presumably more realistic. We also show the pertinent P--waves and
discuss polarization observables. We then turn to some predictions for
neutral pion production off neutrons, with the LECs estimated via
resonance exchange.
\goodbreak \bigskip
\noindent{\bf V.1. A REALISTIC TWO--PARAMETER MODEL FOR $E_{0+} (\omega)$}
\medskip
Tree diagrams and resonance exchanges lead to an almost
constant Re~$E_{0+}$ in
the threshold region, 144.7 MeV $\le E_\gamma < 160$ MeV.
 The unitarity corrections due to the opening of
the $\pi^+ n$ channel at 6.8 MeV above threshold
 have a square--root behaviour as discussed in
section 4.1. This claim is further substantiated by the fact that in
the isospin limit one finds essentially no energy dependence in
Re~$E_{0+}$ [3] but only after inclusion of the pion mass difference
(and, to a lesser extent, the proton--neutron mass difference) a
strong energy dependence develops [19].
To a good approximation, we can therefore parametrize the
 electric dipole amplitude in the threshold region as
$$ E_{0+}(\omega) = -a - b \, \sqrt{1 - (\omega / \omega_c)^2}
\eqno(5.1)$$
which immediately leads to a square--root type behaviour for
Im~$E_{0+}$,
$${\rm Im} \,E_{0+}(\omega) = b \, \sqrt{(\omega / \omega_c)^2 - 1} \,
\Theta(\omega - \omega_c )  \eqno(5.2)$$
which means  that the coefficient $b$ is constrained by the
Fermi--Watson theorem (cf. section 4.5).
We now fit the Mainz data with this form for $E_{0+}$ and
the P--waves as given by the chiral expansion, i.e. we have three
parameters, namely $a$, $b$ and $b_P$, to fit 126 data points (total
and differential cross sections). We find using standard minimization
procedures
$$\eqalign{
 a & = (0.28 \pm 0.07) \cdot 10^{-3}/M_{\pi^+} , \cr
b & = (4.62 \pm 0.49) \cdot 10^{-3}/M_{\pi^+} , \cr
b_P & = (15.64 \pm 0.25) \, {\rm GeV}^{-3} \, . \cr} \eqno(5.3)$$
Several remarks on these numbers are in order. First, the value for
$b_P$ is close to the static $\Delta$ exchange estimate eq.(4.10),
i.e. the multipole $P_3$ is completely dominated by $\Delta$ exchange
even close to threshold.\footnote{$^{6)}$}{The sign of the multipole
$P_3$ is determined from existing multipole analyses at somewhat higher
energies, see e.g. refs.[24]. Of course, only in the
polarization observables to be discussed later this sign plays a role.}
 Second, the fitted value for $b$ is somewhat
larger than what one would get from the Fermi--Watson theorem, $b_{FW}
= 3.7$ (in canonical units of $10^{-3}/M_{\pi^+}$).
The source of this discrepancy is
two--fold. First, in the derivation of Im~$E_{0+}$ from the Fermi--Watson
theorem we assumed exact isospin symmetry and made the further approximation
that the phase shift is simply the product of the scattering length
times the momentum. The result obtained was, however, applied to a
situation involving some isospin breaking.
Second, if the remeasured threshold data
lead to somewhat smaller values of $E_{0+}$ in the threshold region,
 this difference of 20$\%$  would diminish.
Furthermore, the rather simple but physically motivated
form eqs.(5.1,5.2) leads to a good fit with a $\chi^2$/datum of 1.89. At
the respective thresholds, this gives
$$ {\rm Re} \,E_{0+}(\omega_0) = -1.52  \cdot 10^{-3}/M_{\pi^+} \,
\quad  {\rm Re} \,E_{0+}(\omega_c) = -0.28  \cdot 10^{-3}/M_{\pi^+}
\eqno(5.4)$$
which translates into a difference of $\delta E_{0+} = E_{0+}(\omega_c)
 - E_{0+} (\omega_0) = 1.24 \cdot 10^{-3}/M_{\pi^+}$
between the two thresholds. Due to  the reflection properties of
(5.1),
this also means that the imaginary part of $E_{0+}$ at
$\omega_R = 2 \omega_c - \omega_0 = 145.25$ MeV (equivalent to
$E_\gamma^R = 158.27$ MeV) should be
$${\rm Im} \,E_{0+}(\omega_R) = 1.24 \cdot 10^{-3}/M_{\pi^+} \quad .
 \eqno(5.5)$$
Stated differently,  a strong variation of the real part between the
$\pi^0 p$ and the $\pi^+ n$ thresholds reflects itself in a rapid growth
of the imaginary part and vice versa. This stringent constraint
rooted in dispersion theory has not yet been
discussed in the various examinations of the energy--dependence of the
electric dipole amplitude as given by the Mainz data. If one uses a
square--root behaviour of the imaginary part in the threshold region,
the value of Bergstrom [9], Im~$E_{0+}(180$ MeV) $ = 2.0$ translates into
Im~$E_{0+} (\omega_R) = 1.0$ and the four values below 182 MeV given by
M\"ullensiefen [20] lead to Im~$E_{0+} (\omega_R) = (1.0 \pm 0.1)$
(all in canonical units). These numbers are consistent with a direct
calculation of Im~$E_{0+} (\omega_R) = 3.7 \cdot \sqrt{\omega_R^2 /
\omega_c^2 - 1} = 1.0$, and they  indicate that the variation of
Re~$E_{0+}$ between the $\pi^0 p$ and the $\pi^+ n$ thresholds is
indeed less strong as commonly believed. We furthermore stress that
the relative smallness of the parameter $a = 0.3$ indicates that
there have to be large corrections to the Born result of $2.3$ (in
canonical units). Only with the inclusion of loop diagrams, here the
triangle graph and its crossed partner, it is possible to understand
such large corrections to $a$. This can be considered a success of
CHPT. A last important point is the
following. In the approximation (5.1), Re~$E_{0+} (\omega )$ is
strictly constant for $\omega > \omega_c$. At present, the data are
not accurate enough to clearly differentiate between a constant or
slowly varying energy dependence above $\omega_c$. We have therefore
added a linear term of the type $c \,(1 - \omega / \omega_c)$ to
eq.(5.1) and redone the fitting. The values for $a$ and $b$ are
somewhat changed leading to
$ {\rm Re} \,E_{0+}(\omega_0) = -1.57  \cdot 10^{-3}/M_{\pi^+}$ and
${\rm Re} \,E_{0+}(\omega_c) = -0.43  \cdot 10^{-3}/M_{\pi^+}$, not
very different form eq.(5.4) with a comparable $\chi^2$/datum of $1.89$.
However, the one--$\sigma$ uncertainties on
$a$ and $b$ are considerably larger for this type of fit.
This means that a reshuffling
between the linear and the square--root term is possible
(using the existing data). We will come
back to this when we discuss the fit with the counter terms
proportional to $a_1 (\lambda)$ and $a_2 (\lambda)$. We end this
section by stressing again our believe that the pertinent ingredients
 of the threshold behaviour of the electric dipole amplitude are
indeed given by the form eqs.(5.1,5.2).
\goodbreak \bigskip
\noindent{\bf V.2. CHPT RESULTS FOR $\gamma p \to \pi^0 p$}
\medskip
We now turn to the discussion of the results making full use of the
formalism outlined in section 4. First, we consider $a_1 (\lambda)$,
 $a_2 (\lambda)$ and $b_P$ as completely unconstrained parameters and
use the best fit to the Mainz total and differential cross section
data  to determine their values. This will
be called the ``free fit'' in what follows. Second, we vary these
coefficients within the bounds given by the resonance exchange
picture as discussed in section 4.3. This means in particular that we
vary the off--shell parameters $X$, $Y$ and $Z$ (for fixed $g_1 = g_2
= 5$). In principle, one should also vary the vector meson couplings
within some bounds but as discussed before, the uncertainty with
respect to the $\Delta$ parameters is by far larger and we thus use a fixed
vector meson contribution to the various LECs. We will, however, also
present a fit in which $g_1$ and $g_2$ are allowed to vary within
their bounds. We already note here that the results are essentially
indistinguishable from the fit with $g_1 = g_2 = 5$.
This procedure will be coined the  ``resonance fit''.
With $M_{\pi^+} = 140.11$ MeV, the $\pi^+ n$ threshold
is located at its physical value, $E_\gamma = 151.43$ MeV.

First, we show results for the free fit. We find
$$\eqalign{a_1(m) & = (-55.45 \pm 3.34) \, {\rm GeV}^{-4} \, , \cr
a_2(m) & = (58.15 \pm 3.14) \, {\rm GeV}^{-4} \, , \cr
b_P & = (15.80 \pm 0.23) \, {\rm GeV}^{-3} \, . \cr } \eqno(5.6)$$
Notice that the value for $b_P$ is in excellent agreement with the
resonance exchange estimate,
$b_P^\Delta + b_P^V = (12.3+3.1)$ GeV$^{-3}$ = $15.4$ GeV$^{-3}$,
using eqs.(4.7,4.10). Such a value is essentially a consequence
of the bell-shaped differential cross sections for $E_\gamma > 150$
MeV. The result of the free fit for the differential cross sections is
shown in Fig.2 by the solid line (the fit to the data at $E_\gamma =
156.1$ MeV is not exhibited)
and similarly in Fig.3 for the total
cross section.\footnote{$^{7)}$}{Notice that the few Saclay data were not
used in the fitting procedure. Including them would not alter any of
our conclusions.} This fit has a $\chi^2$/datum of $1.88$. If one
multiplies the values of $a_1 (m)$ and $a_2 (m)$ by $M_\pi^3$, one
notices indeed that their individual contributions to $E_{0+}$ at
threshold are of the order $\pm 5.6 \cdot 10^{-3}/M_{\pi^+}$ which is
considerably larger than their sum. The corresponding real part of the
electric dipole amplitude is shown in fig.4a, with
$$ {\rm Re} \,E_{0+}(\omega_0) = -1.56  \cdot 10^{-3}/M_{\pi^+} \,
\quad  {\rm Re} \,E_{0+}(\omega_c) = -0.32  \cdot 10^{-3}/M_{\pi^+}
\eqno(5.7)$$
not very different from the constant plus square--root fit discussed
in the previous section. However, after the $\pi^+ n$ threshold,
Re~$E_{0+} (\omega )$ rises in contrast to the two--parameter
model. We believe that this is an artefact of the strong energy
dependence from the polynomial contact terms proportional to $a_{1,2}$
with their large coefficients. This is reminiscent of the fit we discussed
before when we added a linearly growing term to eq.(5.1). We interpret
the unnaturally large values of $a_1 (m)$ and $a_2 (m)$ as a signal of
the importance of higher loop effects not accounted for by our order $q^4$
calculation. After all, from the discussion of the unitarity
corrections leading to the cusp effect it is rather obvious that one
can not expect a strong energy dependence due to some polynomial terms
(in $\omega$). This is exactly what happens in this free fit - the
unconstrained parameters try to make up for some higher order
effects. Clearly, the situation would be much more satisfactory if
one could determine the LECs $a_{1,2}$ from some other reaction.
The imaginary
part of the electric dipole amplitude (cf. fig.4b) is not affected
by  such uncertainties. It shows the expected square--root type rise and stays
below $0.7 \cdot 10^{-3}/M_{\pi^+}$ for $E_\gamma < 160$ MeV as
already elaborated on in section 4.5. In Fig.5, we exhibit the
conventional P--wave mulitpoles $M_{1+}$, $M_{1-}$ and $E_{1+}$. These
show the empirically expected pattern $M_{1+} > -M_{1-} \gg E_{1+}$ but
still $P_1 \simeq -P_2$ and the $\Delta$ is most visible in $P_3$. We
note that the P--waves are  improved compared to the
relativistic ${\cal O}(q^3)$ calculation [3]. It is furthermore important
to stress that the imaginary parts of these P--waves are tiny
(they increase with $|\vec q \,|^3$) and that
consequently the cusp at the $\pi^+ n$ threshold is not visible. Isospin
breaking effects are also small, typically of the size
$((M_{\pi^+} -M_{\pi^0})/M_{\pi^+})^{3/2} \sim 0.6 \%$.
The energy-dependence of $P_{1,2} /|\vec q \,|$ in the threshold
region $\omega_0 \le \omega \le \omega_R$ is {\it very} weak,
i.e. these reduced P--waves stay constant on the level of 2$\%$.
These observations are at the heart of the
usefulness of the P-wave LETs, eqs.(4.14).

Before discussing the polarization observables, let us consider the
resonance fit. First, we keep $g_1 = g_2 =5$
fixed.\footnote{$^{8)}$}{Note that the empirical width $\Gamma (\Delta
\to N \gamma )$ demands $g_1 \simeq 5$.} We find as best values with a
$\chi^2$/datum of $2.02$
$$ X = 2.24 \pm 1.87 \, , \, \, Y = 0.13 \pm 0.52 \, , \, \, Z = 0.28
\pm 0.75 \, \, \, . \eqno(5.8)$$
This corresponds to $a_1^\Delta (m) = 1.3$ GeV$^{-4}$, $a_2^\Delta (m) = 2.7$
GeV$^{-4}$ and $b_P = 15.9$ GeV$^{-3}$. As already stated, the
magnitude of $a_{1,2}(m)$ is considerably smaller as in the case of
the free fit for the reasons discussed.
The corresponding differential and total cross sections are shown in
Figs.1 and 2 as the dashed lines. They are very similar to the free
fit, the sole exception being the first two MeV above the $\pi^0 p$
threshold. The resonance fit leads to a smaller $E_{0+, {\rm thr}}$
and weaker energy dependence as shown in Fig.4a. Specifically, we have
$$ {\rm Re} \,E_{0+}(\omega_0) = -1.16  \cdot 10^{-3}/M_{\pi^+} \,
\quad  {\rm Re} \,E_{0+}(\omega_c) = -0.43  \cdot 10^{-3}/M_{\pi^+}
\eqno(5.9)$$
This means that the cusp is less pronounced. We point out, however,
that the energy--dependence of $E_{0+}$ for the resonance fit follows
closely the generic form constant plus square root, eq.(5.1) (with
somewhat different values for $a$ and $b$). The large one--$\sigma$
uncertainties on $X$, $Y$ and $Z$ signal that the presently available
data are not yet accurate enough to pin down the electric dipole
amplitude very tightly. In the resonance fit, Re~$E_{0+} (\omega >
\omega_c)$ stays flat as does the two--parameter model discussed
above. Consequently, the relation between the imaginary part
at $\omega_R$ and the difference in the real part between $\omega_0$
and $\omega_c$ is fulfilled (cf fig.4b). The P--waves are essentially
the same as for the free fit.
If we relax the condition that $g_1 = g_2 =5$, we find a
very similar fit with the following values: $X = 0.55 \pm 0.32$, $Y =
0.65 \pm 0.02$, $Z = 0.24 \pm 0.01$, $g_1 = 3.94 \pm 0.06$ and $g_2 =
4.49 \pm 4.24$. Since $g_2$ enters only the S--wave, its value is
determined within large uncertainties. For example, this 5--parameter
fit leads to $E_{0+,{\rm thr}} = -1.17$, only marginally different
from the 3--parameter fit.

We now turn to the polarization observables defined in eq.(2.7). In
Fig.6, we show $\Sigma (\theta)$, $T(\theta)$ and $P(\theta)$ for
$E_\gamma = 153.7$ MeV. Since none of these is sensitive to
Re~$E_{0+}$, the predictions based on the free fit and on the
resonance fits are essentially  the same (compare the solid to the
dashed lines in Fig.6). From the size of the effect we conclude that
the target asymmetry $T(\theta)$ is best suited to pin down the
imaginary part of the electric dipole amplitude. The photon asymmetry
$\Sigma$ is very sensitive to the ratio of the P-wave multipoles
$E_{1+} /M_{1-}$ (for fixed $M_{1+}$).\footnote{$^{9)}$}{This was pointed
out to us by R. Beck, see also ref.[21].} However, note that our
analysis gives a very small $E_{1+}$ in the threshold region so that
the sensitivity discussed in ref.[21] is presumably overestimated.
\goodbreak \bigskip
\noindent{\bf V.3. CHPT PREDICTIONS FOR $\gamma n \to \pi^0 n$}
\medskip
We now discuss predictions for neutral pion production off the neutron
in the threshold region. To fix the low--energy constants, we use the
resonance exchange values discussed in section 4.3 (since no data to
fit exist). Also, we set $m = m_n = 939.57$ MeV. The threshold of this
reaction is at $E_{\gamma}^{\rm thr} = 144.66$ MeV and the $p \pi^-$
channel opens at $E_\gamma^c = 148.46$ MeV, i.e. at $\omega_c = 137.86$
MeV, and we choose the charged pion mass accordingly to account for the
proper location of the second threshold.

The resulting total cross section from threshold up to $E_\gamma =
160$ MeV is shown in fig.7 and four corresponding differential cross
sections in fig.8. We note that $\sigma_{\rm tot}$ rises quicker than
in the case of the proton, this is due to the dual effects of (i) the
larger (in magnitude) electric dipole amplitude,
$E_{0+}^{\pi^0 n}$ changes from 2.13 to 2.77 between threshold and
$E_\gamma = 160$ MeV (in canonical units), and (ii) the even closer
 proximitiy of the first open channel. One also notices the cusp effect.
The differential cross sections are strongly peaked
in forward direction. This can be traced back to the large and positive value
of Re~$E_{0+}$, implying a large and positive angular coefficient
 $B$ (eqs.(2.4,2.5)). This is very different to the case of the
proton. At $\omega_c$, we find $E_{0+}^{\pi^0 n} = 2.79$, i.e the cusp in the
electric dipole is of similar size as in the proton case.
The polarization observables
shown in fig.9 (for $E_\gamma = 153.7$ MeV) are strongly enhanced in  backward
direction which is due to the forward peaked $d\sigma / d\Omega_{\rm cm}$.
 It is obvious that these rather distintive features
should be tested experimentally.
\goodbreak \bigskip
\noindent{\bf VI. SUMMARY, CONCLUSIONS AND OUTLOOK}
\medskip
In this paper, we have used heavy baryon chiral perturbation theory to
study the reactions $\gamma p \to \pi^0 p$ and $\gamma n \to \pi^0 n$
in the threshold region. This is a continuation and improvement on the
calculations making use of relativistic baryon CHPT reported in refs.[3,19].
The pertinent results of this study can be summarized as follows
(preliminary results were presented in ref.[22]):
\medskip
\item{$\bullet$} In the threshold region, one can restrict oneself to
the inclusion of S-- and P--wave multipoles as defined in eq.(2.1). We
have calculated the electric dipole amplitude to order $q^4$ and the
P--waves $P_{1,2,3}$ to order $q^3$. Besides the Born and loop
contributions, we have one finite counter term contribution to $P_3$
and two scale--dependent ones to $E_{0+}$. These counter terms can
either be determined by a best fit (the so--called ``free fit'') to the
threshold data (total and differential cross sections) or estimated from
resonance exchange (``resonance fit''). In the latter case, the
dominant contributions come from the $\Delta (1232)$ as well as the
vector mesons $\rho$ and $\omega$. In both cases we get a good fit to
the existing data. However, for the free fit, the two S--wave
low--energy constants are unnaturally large and of opposite sign. This
signals the importance of higher loop effects not yet accounted for.
The P--wave low--energy constant is essentially given by
$\Delta$--exchange and takes a value expected from the static isobar
model.
\smallskip
\item{$\bullet$} We have considered the low--energy theorem for
$E_{0+}^{\pi^0 p}$ and shown that the convergence in $\mu = M_\pi / m$
is indeed very slow (as conjectured in ref.[3]), compare eq.(4.13).
We conclude  that this multipole is not a good testing ground for
the chiral dynamics of QCD. In case of the neutron, the large
contribution of order $\mu^2$  from the triangle diagram and
its crossed partner appears already at leading order and thus the
convergence for $E_{0+}^{\pi^0 n}$ is much better. From our
calculation, one expects that $|E_{0+}^{\pi^0 n}| > |E_{0+}^{\pi^0
p}|$.
\smallskip
\item{$\bullet$} We have derived novel low--energy theorems for the
P--waves $P_1$ and $P_2$ as given in eqs.(4.14,4.15). These are quickly
converging expansions in $\mu$ and they should be used to constrain
the data analysis. In particular, the small difference in the
magnitudes of $P_1$ and $P_2$ indicates a small but non--vanishing
$E_{1+}$ multipole.
\smallskip
\item{$\bullet$} We have presented a simple but realistic
two--parameter model for the energy depedendence of the electric
dipole amplitude in the threshold region, cf. eqs.(5.1,5.2). The
parameter $b$ is closely related to the strength of the imaginary part of
$E_{0+}$ by the Fermi--Watson theorem. In fact, the energy dependence
of the real part should reflect itself by a similar rise in the
imaginary part above the $\pi^+ n$ threshold. This points towards the
importance of an independent determination of the imaginary part at
$E_\gamma^R = 158.3 $ MeV. \smallskip
\item{$\bullet$} We have discussed the polarization obervables
$\Sigma( \theta), P(\theta)$ and $T(\theta)$ and shown that  the
target asymmetry $T$ seems to be best suited to determine Im~$E_{0+}$
for the neutral pion production off protons. \smallskip
\item{$\bullet$} Finally, we have given predictions for the total and
differential cross sections as well as for polarization observables
for the reation $\gamma n \to \pi^0 n$ (with the low--energy constants
 determined from resonance exchange). These should be determined
experimentally since they serve as a further test of the chiral
dynamics of QCD. \bigskip
Where do we go from here? It is imperative to improve upon the S--wave
on the theoretical side
by either a two--loop calculation or a dispersive representation
constrained by CHPT and on the experimental side by more accurate
determinations of the total and differential cross section close to threshold.
Only with very accurate data one is able to test the proposed constant  plus
square--root form (5.1) for the S--wave constrained by the Fermi--Watson
theorem and the P--wave LETs.
Furthermore, a new look at the corrections to the
low--energy theorems for charged pion photoproduction seems to be
required from the new data on $\pi^-$ production [23]. We hope to come
back to these topics in the future.

\vfill \eject
\noindent{\bf APPENDIX: EXPRESSIONS FOR S-- AND P--WAVE MULTIPOLES}
\medskip
Here, we will give explicit analytical expressions for the S-- and P--wave
multipoles $E_{0+}, \, P_1,\, P_2,\, P_3$ of neutral pion photoproduction from
protons and neutrons. The formulae are given in the isospin limit using only
one pion mass $M_\pi$. In order to account for the branch point and unitarity
cusp above the physical threshold, the value of $M_\pi$ has to be chosen
appropriately, $M_{\pi^+} = \omega_c$, as explained in the text.
\noindent
The Born terms for the proton read:
$$\eqalign{
E_{0+}^{\rm Born}(\omega) &
 = -{e g_{\pi N} \over 24 \pi m^2} \biggl\{ {M_\pi^2
\over \omega} + 2 \omega \biggr\}  + {e g_{\pi N} \over 48 \pi m^3} \biggl\{ 6
\omega^2 + 4 M_\pi^2 - {M_\pi^4 \over \omega^2} + \kappa_p ( 4 \omega^2 -
M_\pi^2) \biggr\} \cr & +{e g_{\pi N} \over 960 \pi m^4} \biggl\{-10  {M_\pi^6
\over \omega^3} + 13 {M_\pi^4 \over \omega} - 156 \omega M_\pi^2 - 72 \omega^3
-10 \kappa_p \omega ( 4\omega^2 + 5 M_\pi^2)\biggr\} \cr } \eqno(A.1)$$
$$P_1^{\rm Born}(\omega) = {e g_{\pi N} |\,\vec q\,|  \over 8\pi m^2} \biggl\{
1 - {6 \omega \over 5m} +  {M_\pi^2 \over 5m\omega} + \kappa_p \biggl(1 -
{\omega \over 2m} \biggr) \biggr\} \eqno(A.2)$$
$$P_2^{\rm Born}(\omega) = {e g_{\pi N}|\,\vec q\,|\over 8\pi m^2} \biggl\{ -1
+ {13 \omega \over 10m} +  {M_\pi^2 \over 5m\omega} + \kappa_p \biggl(-1 +
{\omega \over 2m} \biggr)  \biggr\} \eqno(A.3)$$
$$P_3^{\rm Born}(\omega) = {e g_{\pi N} |\,\vec q\,|\,\omega  \over 16\pi m^3}
\eqno(A.4)$$
The Born terms for the neutron are:
$$E_{0+}^{\rm Born}(\omega) = {e g_{\pi N} \kappa_n\over 48 \pi m^3} \bigl\{
M_\pi^2- 4 \omega^2  \bigr\} +{e g_{\pi N} \kappa_n \over 96 \pi m^4}
\omega \,\bigl\{4\omega^2 + 5 M_\pi^2\bigr\} \eqno(A.5)$$
$$P_1^{\rm Born}(\omega) = - P_2^{\rm Born}(\omega) = {e g_{\pi N} |\,\vec q\,|
\over 8\pi m^2}\,\kappa_n\, \biggl\{-1 +{\omega \over 2m} \biggr\}
\eqno(A.6)$$
$$P_3^{\rm Born}(\omega) = 0 \eqno(A.7)$$
The loop contributions  for the proton and the neutron differ only by
a few numerical coefficients:
$$E_{0+}^{\rm q^3-loop}  (\omega)
= {eg_A \over 64 \pi^2 F_\pi^3} \biggl\{ M_\pi^2 \arcsin{\omega\over
M_\pi} - \omega \sqrt{M_\pi^2 - \omega^2}\biggr\} \eqno(A.8)$$
$$\eqalign{ E_{0+}^{\rm q^4-loop} & (\omega)  =
  {e g_A \over 128 \pi^3
m F_\pi^3} \biggl\{{2\over3}(11+4\tau^3)\omega^3 \ln{M_\pi\over\lambda} +
{8\over 9}(1-\tau^3)\omega^3- {4\over 3}(2+\tau^3) \omega M_\pi^2 \cr & +
\sqrt{M_\pi^2-\omega^2} \biggl( \pi M_\pi^2 +{2\over 3} [(11+4\tau^3)\omega^2
+(1+2\tau^3) M_\pi^2] \arcsin
{\omega \over M_\pi} \biggr) \cr & - \omega M_\pi^2 \biggl( 2\pi + \arcsin
{\omega\over M_\pi} \biggr) \arcsin{\omega\over M_\pi} \cr & +  {g_A^2 \over 9}
 \biggl[ 2\omega (7\omega^2 + 5 M_\pi^2)
 \ln{M_\pi\over\lambda}  + {2\over3} \omega^3 - {20\over 3} \omega M_\pi^2 -5
{ M_\pi^4\over \omega} + \pi M_\pi(12 \omega^2 - M_\pi^2 + 4 {M_\pi^4 \over
 \omega^2} ) \cr & + \sqrt{M_\pi^2-\omega^2} \biggl( \pi ( 8 \omega^2 -13
 M_\pi^2 - 4 {M_\pi^4 \over \omega^2} )
+ 2(7\omega^2 +M_\pi^2 +{ M_\pi^4 \over \omega^2}
 ) \arcsin{\omega \over M_\pi} \biggr) \cr & + 3 \pi  M_\pi^2(
{M_\pi^2 \over
 \omega} - 4 \omega)  \arcsin{\omega\over M_\pi}+ 3 M_\pi^2 ( {M_\pi^4 \over
 \omega^3} + {3 M_\pi^2 \over \omega} - \omega)\arcsin^2 {\omega\over M_\pi}
\biggr] \biggr\} \cr} \eqno(A.9)$$
$$P_1^{\rm loop}(\omega) = {eg_A^3 \, |\,\vec q\,| \over 64 \pi^2
F_\pi^3}\biggl\{ {2\over 3 \omega^2} \bigl[ M_\pi^3 - (M_\pi^2-\omega^2)^{3/2}
\bigr] + M_\pi - \sqrt{M_\pi^2 -\omega^2} - {M_\pi^2 \over \omega}
\arcsin{\omega\over M_\pi} \biggr\} \eqno(A.10)$$
$$P_2^{\rm loop}(\omega) = {eg_A^3 \, |\,\vec q\,| \over 64 \pi^2
F_\pi^3}\biggl\{ {2\over 3 \omega^2} \bigl[ M_\pi^3 - (M_\pi^2-\omega^2)^{3/2}
\bigr] - M_\pi \biggr\} \eqno(A.11)$$
$$P_3^{\rm loop}(\omega) = 0 \eqno(A.12)$$
The analytic continuation of these expressions above the branch point $\omega
= M_\pi$ is obtained through the following substitutions:
$$\sqrt{M_\pi^2-\omega^2} \to -i \, \sqrt{\omega^2-M_\pi^2}\, , \qquad
\arcsin{\omega\over M_\pi} \to {\pi \over 2} + i \, \ln
{\omega+\sqrt{\omega^2-M_\pi^2}  \over M_\pi} \eqno(A.13)$$
The counterterm contributions for the protons and neutrons are:
$$E_{0+}^{\rm ct}(\omega) = e\, a_2^{p,n}(\lambda) \,\omega^3 + e\, a_1^{p,n}
(\lambda)\,\omega \,M_\pi^2 \eqno(A.14)$$
$$P_1^{\rm ct}(\omega) = P_2^{\rm ct}(\omega) = 0 \eqno(A.15)$$
$$P_3^{\rm ct}(\omega) = e \,b_P^{p,n} \,\omega \,|\,\vec q\,| \eqno(A.16)$$
\vfill \eject
\noindent{\bf REFERENCES} \bigskip
\item{1.}E. Mazzucato {\it et al.},
 {\it Phys. Rev. Lett.\/} {\bf 57} (1986) 3144;

R. Beck {\it et al.}, {\it Phys. Rev. Lett.\/} {\bf 65} (1990) 1841.
\smallskip
\item{2.}T. P. Welch et al., {\it Phys. Rev. Lett.\/}
{\bf 69} (1992) 2761.
\smallskip
\item{3.}V. Bernard, J. Gasser, N. Kaiser and Ulf-G. Mei{\ss}ner,
{\it Phys. Lett.\/} {\bf B268} (1991) 219;
V. Bernard, N. Kaiser, and Ulf-G. Mei{\ss}ner, {\it Nucl. Phys.\/}
{\bf B383} (1992) 442.
\smallskip
\item{4.}J. Gasser, M.E. Sainio and A. {$\rm \check S$}varc,
{\it Nucl. Phys.\/} {\bf B307} (1988) 779.
\smallskip
\item{5.}E. Jenkins and A.V. Manohar, {\it Phys. Lett.\/} {\bf B255} (1991)
558.
\smallskip
\item{6.}V. Bernard, N. Kaiser, J. Kambor
and Ulf-G. Mei{\ss}ner, {\it Nucl. Phys.\/} {\bf B388} (1992) 315.
\smallskip
\item{7.}V. Bernard, N. Kaiser, T.--S. H. Lee and Ulf-G. Mei{\ss}ner,
 {\it Phys. Rep.} {\bf 246} (1994) 315.
\smallskip
\item{8.}A.M. Bernstein, $\pi N$ {\it Newsletter} {\bf 9} (1993) 55.
\smallskip
\item{9.}
A. M. Bernstein and B. R. Holstein,  {\it Comments Nucl. Part. Phys.\/}
{\bf 20} (1991) 197;

J. Bergstrom,  {\it Phys. Rev.\/} {\bf C44} (1991) 1768;

L. Tiator,
''Meson Photo- and Electroproduction'', lecture given at the II TAPS

Workshop, Alicante, 1993;

A.M. Bernstein, private communication.

\smallskip
\item{10.}G. Ecker, {\it Phys. Lett.\/} {\bf B336} (1994) 508.
\smallskip
\item{11.}G. Ecker, J. Gasser, A. Pich and E. de Rafael,
{\it Nucl. Phys.} {\bf B321} (1989) 311;

J.F. Donoghue, C. Ramirez and G. Valencia, {\it Phys. Rev.} {\bf D39} (1989)
1947. \smallskip
\item{12.} Ulf--G. Mei{\ss}ner, ``Aspects of Nucleon
Chiral Perturbation Theory'', talk given at the Workshop on Chiral
Dynamics: Theory and Experiments, MIT, Cambridge, July 1994, preprint
CRN-94/44. \smallskip
\item{13.}V. Bernard, N. Kaiser, Ulf-G. Mei{\ss}ner and A. Schmidt,
{\it Z. Phys.\/} {\bf A348} (1994) 317. \smallskip
\item{14.}M. Benmerrouche, R.M. Davidson and N.C. Mukhopadhyay,
 {\it Phys. Rev.} {\bf C39} (1989) 2339. \smallskip
\item{15.} G. Ecker and Ulf--G. Mei{\ss}ner, ``What is a Low--Energy
Theorem ?'',  preprint CRN-94/52 and UWThPh-1994-33. \smallskip
\item{16.}J. Gasser, H. Leutwyler and M.E. Sainio,
{\it Phys. Lett.\/} {\bf B253} (1991) 252,260.   \smallskip
\item{17.}J. Gasser and Ulf--G. Mei{\ss}ner,
{\it Nucl. Phys.} {\bf B357} (1991) 90. \smallskip
\item{18.}D. Drechsel and L. Tiator,
{\it J. Phys. G: Nucl. Part. Phys.} {\bf 18} (1992) 449. \smallskip
\item{19.}V. Bernard, N. Kaiser,
and Ulf-G. Mei{\ss}ner, $\pi N$ {\it Newsletter\/} {\bf 7} (1992) 62.
\smallskip
\item{20.}A. M\"ullensiefen, {\it Z. Phys.\/} {\bf 211} (1968) 360.
\smallskip
\item{21.}J. Ahrens et al., MAMI proposal A2/7-94,
October 1994.
\item{22.}V. Bernard, ``Threshold Pion Photo- and Electroproduction in
Chiral Perturbation Theory'', talk given at the Workshop on Chiral
Dynamics: Theory and Experiments, MIT, Cambridge, July 1994, preprint
CRN-94/45. \smallskip
\item{23.}M. Kovash, talk given at the Workshop on Chiral
Dynamics: Theory and Experiments, MIT, Cambridge, July 1994;

G. H\"ohler, private communication. \smallskip
\item{24.}F.A. Berends and D.L. Weaver,
{\it Nucl. Phys.\/} {\bf B30} (1971) 575;

F.A. Berends and A. Donnachie, {\it Nucl. Phys.\/} {\bf B84} (1975) 342.
\smallskip
\bigskip \bigskip \bigskip
\noindent{\bf FIGURE CAPTIONS} \bigskip
\item{Fig.1}Feynman diagrams contributing at ${\cal O}(q^3)$. The
triangle (a) and rescattering (b) diagrams give $E_{0+}(\omega )$
while (c) and (d) contribute to the P--wave multipoles $P_1$ and $P_2$.
Crossed graphs are not shown.
\item{Fig.2}Differential cross sections for $\gamma p \to \pi^0
p$. The solid lines refer to the free fit and the dashed ones to the
resonance fit as explained in the text. The data are from Mainz[1].
\smallskip
\item{Fig.3}Total cross section for $\gamma p \to \pi^0 p$. For
notations, see fig.2.
The data are from Mainz (diamonds) and Saclay (squares) [1].
\smallskip
\item{Fig.4}The electric dipole amplitude for $\gamma p \to \pi^0
p$. (a) The real part. For
notations, see fig.2. In addition, the $1\sigma$--bands for the free
fit are indicated by the dotted lines. (b) Imaginary part.
\smallskip
\item{Fig.5}The P--wave multipoles $M_{1+}$, $M_{1-}$ and $E_{1+}$ for
the free fit. \smallskip
\item{Fig.6}The polarization observables $\Sigma (\theta)$, $T(
\theta)$ and $P (\theta)$ for $E_\gamma = 153.7$ MeV. For notations,
see fig.2. \smallskip
\item{Fig.7}Chiral prediction for $\sigma_{\rm tot} (\gamma n \to \pi^0
n)$.  The low--energy constants are estimated from resonance exchange.
\smallskip
\item{Fig.8}Chiral prediction for the differential cross section for
$\gamma n \to \pi^0 n$ at $E_\gamma = 149.1$ (dashed), 151.4 (dotted),
153.7 (solid) and 156.1 (dash-dotted) MeV.
\smallskip
\item{Fig.9}Chiral prediction for the polarization observables $\Sigma,
P, T$ for $\gamma n \to \pi^0 n$ at $E_\gamma = 153.7$ MeV.
\smallskip

\vfill \eject \end